\documentclass[a4paper,11pt]{article}
\usepackage[pdftex]{graphicx}
\title{Cooperation for public goods under uncertainty}
\author{Jeroen Bruggeman and Rudolf Sprik }
\date{\today}
\begin{document}

\maketitle

\begin{abstract} Everyone wants clean air, peace and other public goods but is tempted to freeride on others' efforts. The usual way out of this dilemma is to impose norms, maintain reputations and incentivize individuals to contribute. In situations of high uncertainty, however, such as confrontations of protesters with a dictatorial regime, the usual measures are not feasible, but cooperation can be achieved nevertheless. We use an Ising model with asymmetric spins that represent cooperation and defection to show numerically how public goods can be realized. Under uncertainty, people use the heuristic of conformity. The turmoil of a confrontation causes some individuals to cooperate accidentally, and at a critical level of turmoil, they entail a cascade of cooperation. This critical level is much lower in small networks. \end{abstract}

\section{Introduction}
A main benefit for people living in groups, ranging from families to empires, is that collectively, they can achieve more than the sum of individuals can independently, in particular collective goods \cite{gavrilets15}. Cases in point are infrastructure, health care, defense and education for a country's population. People first have to agree on the public good and the way to realize it, but even if they arrive at a consensus, the provision of these goods is non-obvious because individuals are tempted to freeride on others' efforts. Cooperation is a dilemma \cite{olson65} that usually requires time to solve, during which people develop norms and form a network to transmit information (gossip) about one another, on the basis of which reputations are established that in their turn are used to reward cooperators and punish defectors  \cite{fehr03,nowak05}.    

This scenario is well-covered by the literature, but there are important cases where there is no time to develop it. Examples are disasters where victims need urgent help, protests against dictatorial regimes that prevent critics to organize themselves, and unplanned violent confrontations between groups.  These cases have high uncertainty in common. People then realize that cooperation might yield a valuable outcome but they can't assess their payoffs, among others because they don't know if they will get hurt. Our questions are how cooperation under uncertainty is self-organized, and how this is influenced by participants' network.

The first question has been addressed by critical mass theory \cite{granovetter78,marwell93}, which does not rely on norms, reputations and selective incentives (i.e. rewards and punishments). It argues that if a critical number of actors starts cooperating, the others are won over to join in, thereby making cooperation self-reinforcing. Individuals are assumed to know at each moment how many cooperators there are in total, and to have fairly accurate expectations of their marginal payoff. On the basis thereof they can rationally decide to contribute to the public good or to freeload. In highly uncertain situations, however, these rationality assumptions are unlikely to hold true. Moreover, the theory does not show the effect of network topology, and does not explain the cooperation of the initiative takers before the critical point is reached. We, in contrast, explicate the effect of topology, explain the initiative takers endogenously, and instead of rationality assume one simple heuristic:
under high uncertainty, people become conformists to the majority of their network neighbors \cite{wu14}. This may turn out bad for them in a specific situation but may serve them on average over many occasions \cite{vandenberg18}. Conformism can be observed as behavioral synchronization \cite{mcneill95}.  When people conform to their initially defecting network neighbors, they might eventually cooperate when many others do, but who would start? 

Pending (or starting) violence and disaster are characterized by increasing turmoil, for example in terms of opponents'  threats, insults and violence. Turmoil can be measured by participant's heart rates that indicate their arousal \cite{konvalinka11} or, in an information theoretic manner, by accelerating situational updates to the same effect \cite{johnson16}. Arousal causes ``trembling hands" \cite{dion88} as game theorists say, denoting a chance that some individuals accidentally cooperate, which in turn might entail a cascade of cooperation. An example of turmoil is the mutual provocation of soccer fans of opposing camps to a boiling point when fighting breaks out; their public good is the humiliation of their opponents. To show that turmoil can drive cooperation, we use an Ising model \cite{barrat08,castellano09},  for which we make the behavioral options (spins)  asymmetric, reflecting the asymmetry of defection versus cooperation. This provides a parsimonious explanation of cooperation without the usual mechanisms developed over longer time, and without rationality assumptions.

\section{Model}
At the beginning there is a group of $n$ individuals who may not know one another yet, in an uncertain situation. Each of them has at least visual contact with some others, and identifies with, and is therefore inclined to conform to, others who share an interest in a given public good. The network of visual or verbal ties $A_{ij}$ denoting that $i$ pays attention to $j$ is represented by a row-normalized adjacency matrix with cells $ a_{ij} = A_{ij} / \sum_j A_{ij}$, as in many models of social influence \cite{friedkin11}, hence $ \sum_j a_{ij}$ = 1. Ties are bi-directional but asymmetric, and the network may have multiple disconnected components. An individual can cooperate ($C$) or defect ($D$) with $C > D > 0$, and everybody defects at the beginning. The behavioral variable $S_i$ can take the value $ S_i=C $ or $ S_i=-D $.  Behavior and network are expressed in the Hamiltonian of the Ising model
\begin{equation}
H = - \sum_{i \neq j}^{n} a_{ij} S_i S_j.
\label{eq:ising1}
\end{equation}  

\begin{figure}
\begin{center}
\includegraphics[width=0.8\textwidth]{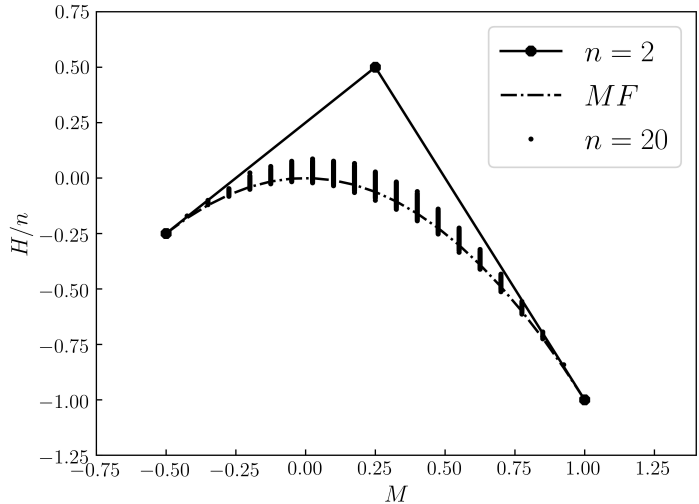}
\end{center}
\caption{Ising model with $ S = \{1,-1/2\} $.  Mean dissatisfaction $H/n$ at different levels of cooperation $M$ in a network with $n=2$, and in a random net with $n=20$ and density = 0.8. In the latter, vertical line segments are composed of dots, each representing a micro state. The dash-dotted line is drawn with the mean field approach (MF). }
\label{fig:Ising}
\end{figure} 

The mean degree of cooperation is described by an order parameter $ M = 1/n \sum_{i = 1}^{n} S_i $.  A cooperator's payoff is a share of the public good at a cost $P_C = r(M + C/n) - C$, with a synergy parameter $r>1$; a defector's payoff is $P_D = r(M - D/n)$ \cite{perc17}. In conventional game theory, $C = 1$ and $D = 0$ \cite{perc17}, but since payoffs are used only comparatively, it does not matter if they are negative here. If people believe that $C \leq D$, this payoff function does not apply and people will not cooperate anyway. If they believe that $C > D$, they may cooperate provided that enough others do. Because under high uncertainty, participants can't assess the synergy of their collective effort, don't know the costs and benefits, and in large groups can't see all cooperators, they can't calculate their payoff. The conformism heuristic then works as a collective lever that minimizes $H$ and can maximize the mean payoff indirectly, even without individuals knowing the payoff function. Individuals' heuristic decision making is implemented computationally by the Metropolis algorithm \cite{barrat08}. Over a large number of Monte Carlo steps, a node $i$ is chosen randomly, its behavior $S_{i}$ is flipped from $D$ to $C$ (or the other way around), and $H$ is compared to $H'$ that has the flip implemented and is otherwise identical to $H$. The flip is accepted if for a random number $ 0 \leq c_r \leq 1$,
\begin{equation}
c_r < exp \bigg(\frac{-(H' - H)}{T} \bigg),
\label{eq:metropolis}
\end{equation}  
where $T$ is the level of turmoil, or temperature in the original model \cite{barrat08}.     
$H/n$ can be regarded as mean dissatisfaction with respect to the conformity heuristic, biased towards cooperation for the public good. The relation between $H/n$ and $M$ is illustrated in Fig.~\ref{fig:Ising} for $S=\{1,-1/2\}$ on a network with $n = 2$, for a random net with $n = 20$ including all micro states (configurations of $C$'s and $D$'s) therein, and in a mean field approach (MF) for comparison. The Metropolis minimization of $H$ starts at the defective state (at the left) and evolves by random steps over all micro states until the lowest value of $H$ is reached at the cooperative state (at the right).

\section{Results}
\begin{figure}
\begin{center}
\includegraphics[width=0.70\textwidth]{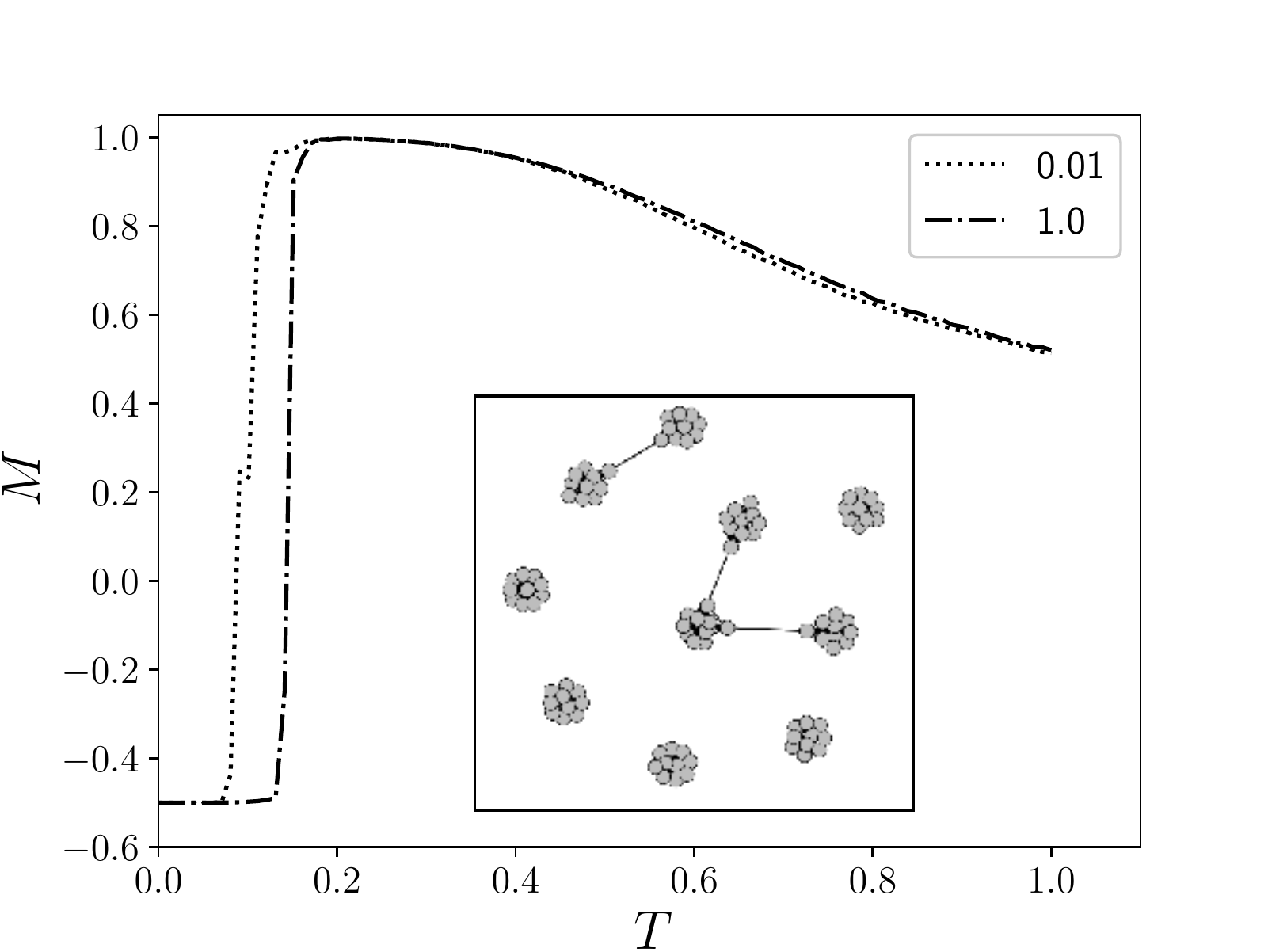}
\end{center}
\caption{Ising model with $ S = \{1,-1/2\} $, showing $M$ with increasing $T$ for a network with $n=100$ and 10 clusters, $n=10$ each. There is a chance $p$ for ties to be rewired. If $p = 0$, the clusters are fully connected cliques, mutually disconnected. The network at $p = 0.01$ is shown at the inset and with a dotted line in the $M-T$ plot.
If $p = 1$ the network becomes essentially a random graph, with a higher $T_c$.}
\label{fig:clustera}
\end{figure}

The relation between cooperation, $M$, and turmoil, $T$, turns out to be qualitatively the same for all networks. At low turmoil, there is no cooperation but it emerges at a critical level $T_{c}$, illustrated in Fig.~\ref{fig:clustera}. Then the turmoil overcomes the ``energy" barrier (the mountain in Fig.~\ref{fig:Ising}), and drives the transition to the cooperative state. For as long as $T < T_c$, the chance that this happens in a finite number of Monte Carlo steps is zero. If $T$ keeps increasing beyond $T_c$, the effect of turmoil becomes progressively more dominant until individuals randomly alternate cooperation and defection. Here, turmoil trumps conformism. If not by this chaos, cooperation will end when the public good is realized, others intervene, the participants get exhausted, or if they get to understand the situation (i.e.~their uncertainty lowers) and start behaving strategically (i.e.~freeride).   

\begin{figure}
\begin{center}
\includegraphics[width=0.70\textwidth]{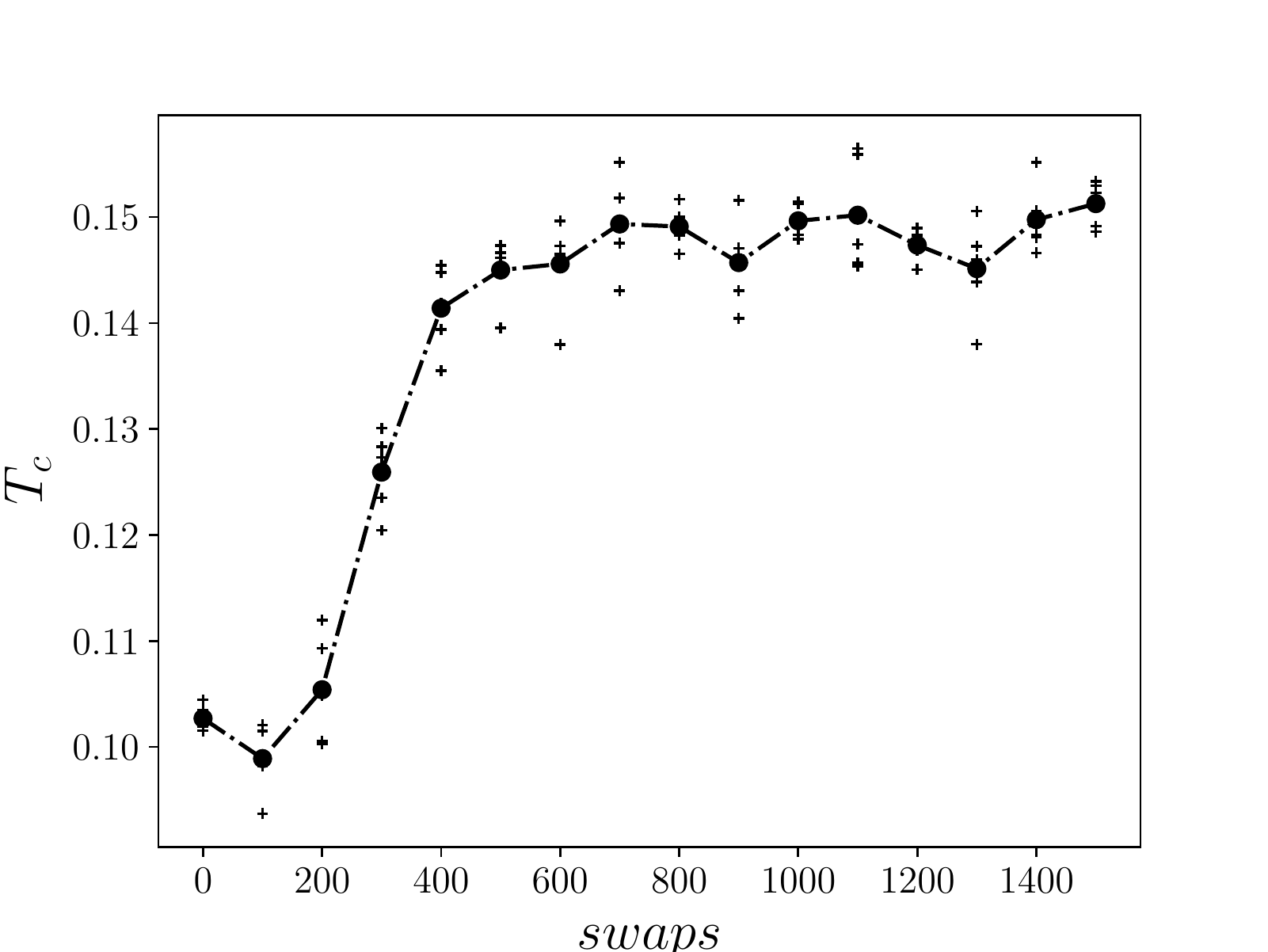}
\end{center}
\caption{Effect of increasing rewiring on the tipping point $T_c$ in a $n = 100$ graph with 10 clusters, $n=10$ each, as in Fig.~\ref{fig:clustera}. Again, $S = \{1,-1/2\} $. The horizontal axis indicates the number of swaps. A swap is generated by randomly choosing three nodes with 1 and 2 connected but not 1 and 3. Then 1's tie to 2 is relayed to 3. The simulation was repeated five times (plus signs) and resulted in an average value indicated by the dots connected by the dash-dotted line.}   
\label{fig:clusterdetail}
\end{figure} 

\begin{figure}
\begin{center}
\includegraphics[width=0.75\textwidth]{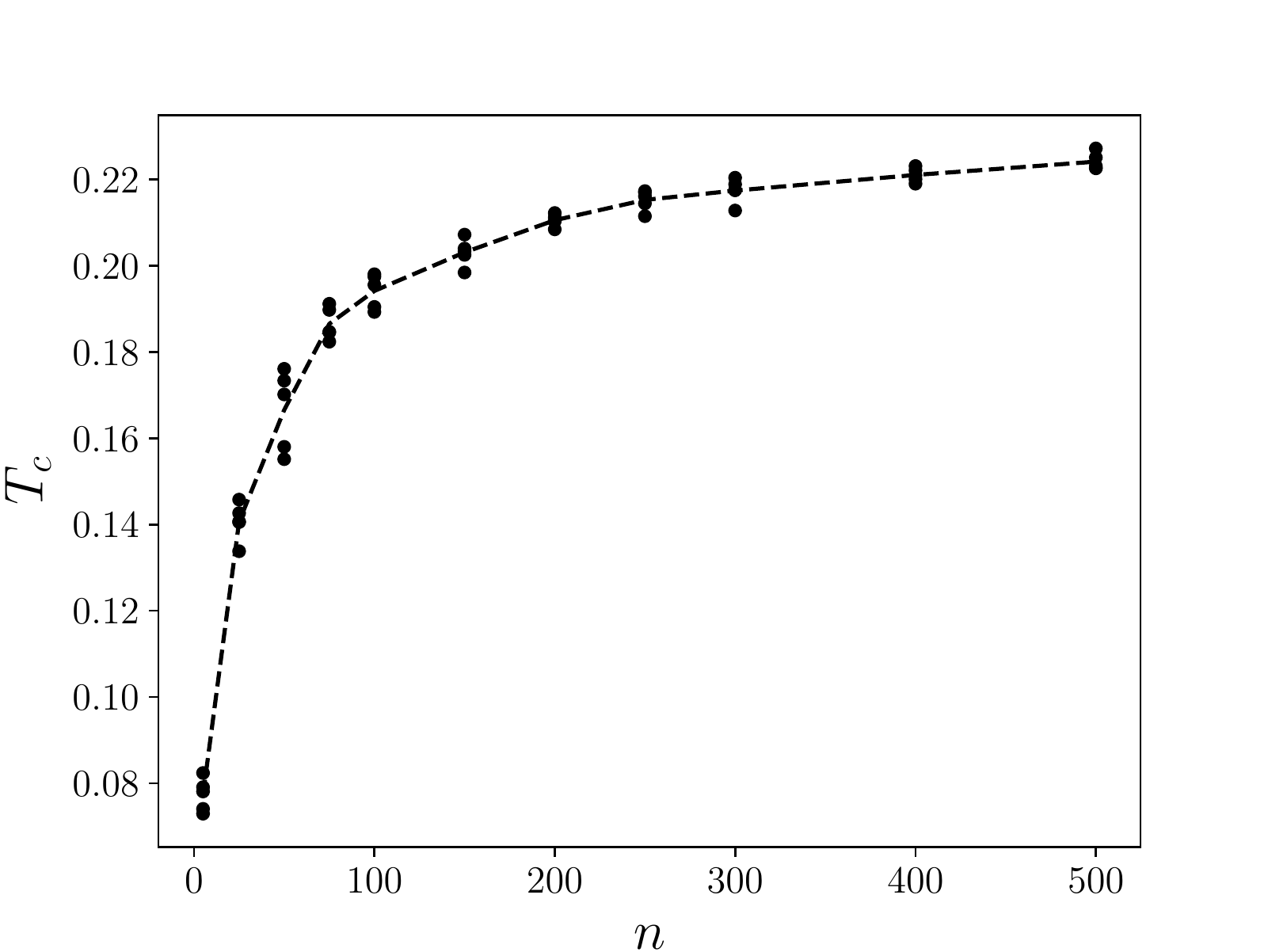}
\end{center}
\caption{Effect of increasing network size $n$ on the tipping point $T_c$ in random networks with density = 0.8 and $ S = \{1,-1/2\} $. The dots show small variation across simulation runs.}
\label{fig:size} 
\end{figure}

Social networks, if large, are clustered and sparse with a skewed degree distribution \cite{newman18}. We therefore examine the effects of clustering, density, degree distribution and size on the tipping point $T_c$, and subsequently zoom in on local variations of $S$. 
As a characteristic example of clustering, we start with ten cliques (fully connected networks) with $n=10$ each (Fig.~\ref{fig:clustera}). When randomly rewiring ties with a probability $p$, the network becomes more random at higher $p$ and thereby less clustered, as in small world networks \cite{newman18}. Decreasing clustering results in increasing $T_c$. 
In Fig.~\ref{fig:clusterdetail} this is demonstrated by starting with a strongly clustered network, and by rewiring the links, which will often form random cross-cluster connections. From the most strongly clustered network onward, the largest effect on $T_c$ is at small numbers of rewired ties. 
The effect of network size on the tipping point is larger than of clustering, though, in particular at the smallest network sizes (Fig.~\ref{fig:size}). $T_c$ is also lower in sparser networks but the effect of density is much weaker than of size and more variable across simulation runs (not shown). When varying the degree distribution between Poisson and power law, $T_c$ does not change at all. 
In sum, changes of $T_c$ are largest at small network size, high clustering, and low density.

To investigate the effect of local variations of $S$, we generalize the values of $S = \{1,-1/2\}$ that we used in the above examples. To this end we rewrite the asymmetric Ising model in a symmetric form by the mapping
\begin{equation}
S = \{C, -D\} \rightarrow \{ S_0 + \Delta, S_0 - \Delta \},
\end{equation}
with a bias $S_0 = (C-D)/2$ and an increment $\Delta = (C+D)/2$. Accordingly, the values chosen in our examples imply $S_0 = 0.25$ and $\Delta = 0.75$. Elsewhere we show that the Hamiltonian, once written in a symmetric form, becomes a conventional Ising model plus a global external field and a locally varying field that depends on the network \cite{brugspin}. We use this generalized form in a mean field analysis, from which we infer that for given $\Delta$, higher $S_0$ (that corresponds to a stronger interest in the public good) results in lower $T_c$ \cite{brugspin}. 


\begin{figure}
\includegraphics[width=0.99\textwidth]{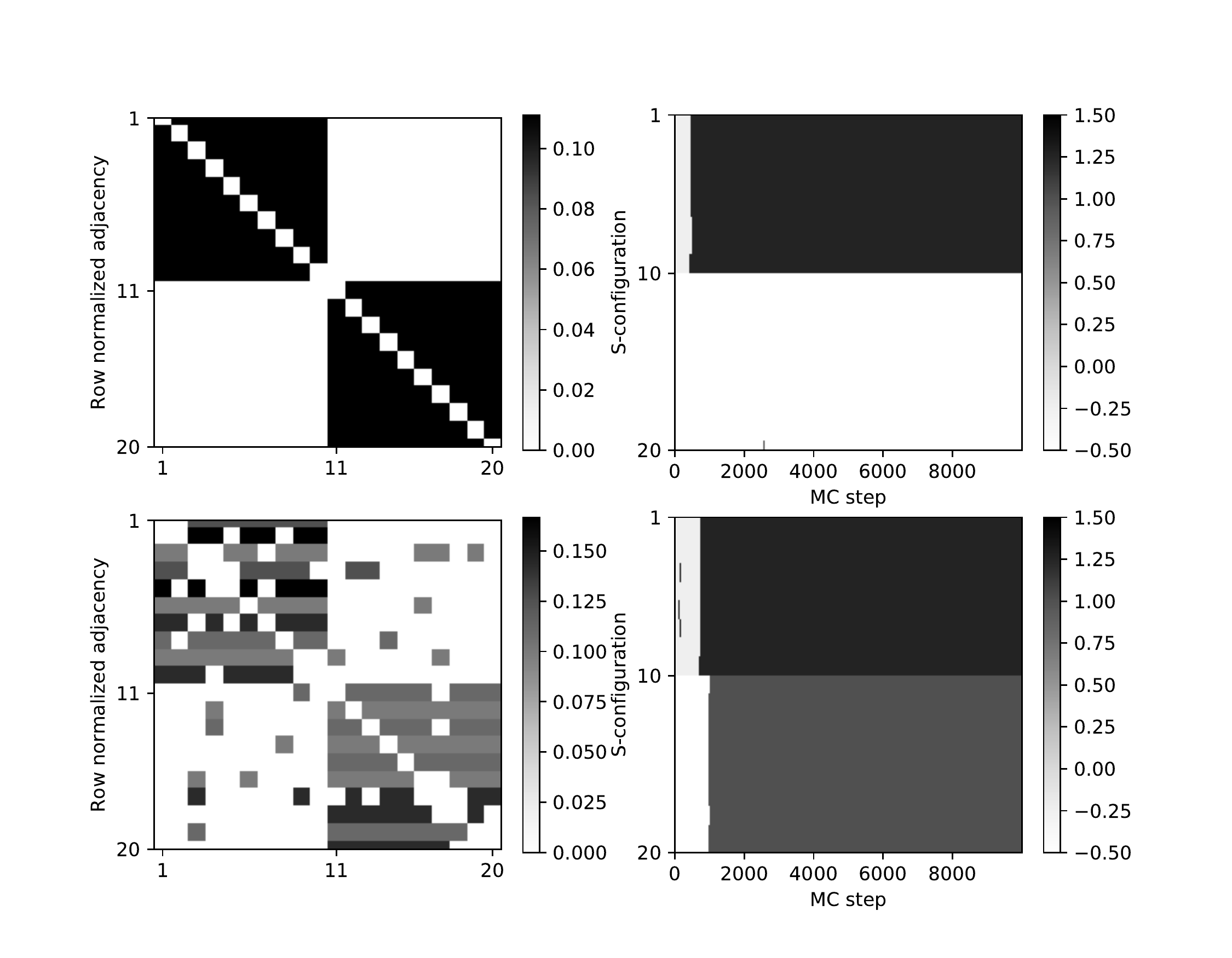}
\caption{Networks with two groups at $T = 0.09$. Top row left: adjacency matrix of two disconnected groups ($p=0, n=10$), with $S_0 = 0.50$ (nodes 1--10) and $S_0 = 0.25$ (nodes 11--20), respectively. The top row right panel has Monte Carlo steps on the horizontal axis. The group with higher $S_0$ (nodes 1--10) cooperates after few Monte Carlo steps whereas the group with lower $S_0$ (11--20) does not cooperate.  Bottom row left: for the same two groups, ties are randomly rewired with a chance $p=0.1$, resulting in connections between the groups that are visible in the adjacency matrix. Bottom row right: the group with higher $S_0$ (1--10) wins over the other group (11--20) to cooperate after few Monte Carlo steps. The different gray tones for the two cooperating subgroups are due to their different $S_0$ values.}
\label{fig:twogroups} 
\end{figure}

We now use different $S_0$ in a network with two clusters, keeping $\Delta = 0.75$.
When $p=0$ the two groups are disconnected, but within-group links can be randomly rewired with a probability $p$ and thereby connect the two groups. In Fig.~\ref{fig:twogroups}, two two-cluster networks ($n=20$) with $p = 0$ and $p = 0.1$, respectively, are illustrated at the top and bottom row; $T = 0.09$ for both. In each network, all members of one cluster (top) have $S_0 = 0.50$ and all members of the other cluster (bottom) have $S_0 = 0.25 $.  
For the network with disconnected groups, the group with lower $S_0$ does not reach its tipping point. When the groups are connected ($p = 0.1$), the group with higher $S_0$ helps the other group to cooperate at lower $T_c$ than it would on its own, whereas the former is somewhat held back by the latter. This result generalizes to small numbers of initiative takers or leaders with higher $S_0$ embedded in a network of many others with lower $S_0$, where the initiative takers reduce $T_c$ for the majority (not shown).




\section{Conclusions}
The model points out that under high uncertainty, the dilemma of cooperation can in principle be solved by situational turmoil among conformists. This solution holds for a broad range of networks, although the critical level of turmoil is markedly lower in small ones, which implies that turmoil-driven cooperation is more likely in small than in large groups. This result is consistent with empirical findings on violent group confrontations, where most of the physical violence is committed by small subgroups \cite{collins08}. Small fighting groups increase overall turmoil, though, which may agitate larger groups to join. The model also shows that there is no need for especially zealous initiative takers (with high $S_0$) to get cooperation started, even though they help to get it started at lower turmoil.

Our contributions are threefold. First, we made an asymmetric Ising model, which has a different phase transition than the widely used symmetric model. Second, we provided computational evidence that the dilemma of cooperation can be solved without complicated and costly mechanisms such as reputations, selective incentives and rationality.  Third, we showed which networks are most conductive to cooperation under high uncertainty.
Future studies could confront the model with empirical data, perhaps even of different species, for example buffalo herd bulls who chase a lion together \cite{estes91}. One could also investigate more systematically the numbers of necessary Monte Carlo steps for different networks and distributions of $S_0$.


\end{document}